\documentstyle[12pt]{article}

\font\twlgot =eufm10 scaled \magstep1 \font\egtgot =eufm8
\font\sevgot =eufm7 \font\twlmsb =msbm10 scaled \magstep1
\font\egtmsb =msbm8 \font\sevmsb =msbm7

\newfam\gotfam
\def\pgot{\fam\gotfam\twlgot}
\textfont\gotfam\twlgot \scriptfont\gotfam\egtgot
\scriptscriptfont\gotfam\sevgot
\def\got{\protect\pgot}
\newfam\msbfam
\textfont\msbfam\twlmsb \scriptfont\msbfam\egtmsb
\scriptscriptfont\msbfam\sevmsb
\def\Bbb{\protect\pBbb}
\def\pBbb{\relax\ifmmode\expandafter\Bb\else\typeout{You cann't use
Bbb in text mode}\fi}
\def\Bb #1{{\fam\msbfam\relax#1}}

\newcommand{\gT}{{\got T}}

\def\thebibliography#1{\bigskip\section*{}\bigskip\list
{$^{\arabic{enumi}}$}{\settowidth\labelwidth{#1}\leftmargin\labelwidth
\advance\leftmargin\labelsep
\usecounter{enumi}}
\def\newblock{\hskip .11em plus .33em minus .07em}
\sloppy\clubpenalty4000\widowpenalty4000 \sfcode`\.=1000\relax}

\def\op#1{\mathop{\fam0 #1}\limits}

\newcommand{\beq}{\begin{equation}}
\newcommand{\eeq}{\end{equation}}
\newcommand{\ben}{\begin{eqnarray}}
\newcommand{\een}{\end{eqnarray}}
\newcommand{\be}{\begin{eqnarray*}}
\newcommand{\ee}{\end{eqnarray*}}
\newcommand{\bea}{\begin{eqalph}}
\newcommand{\eea}{\end{eqalph}}
\newcommand{\cA}{{\cal A}}

\newcommand{\cD}{{\cal D}}

\newcommand{\cE}{{\cal E}}
\newcommand{\cH}{{\cal H}}

\newcommand{\bL}{{\bf L}}

\newcommand{\bT}{{\bf T}}

\newcommand{\la}{\lambda}

\newcommand{\f}{\phi}
\newcommand{\om}{\omega}
\newcommand{\Om}{\Omega}

\newcommand{\g}{\gamma}
\newcommand{\G}{\Gamma}

\newcommand{\vt}{\vartheta}

\newcommand{\lng}{\langle}
\newcommand{\rng}{\rangle}

\newcommand{\w}{\wedge}
\newcommand{\wt}{\widetilde}
\newcommand{\wh}{\widehat}
\newcommand{\ol}{\overline}
\newcommand{\dr}{\partial}

\newcommand{\ot}{\otimes}
\newcommand{\ap}{\approx}

\newcounter{eqalph}
\newcounter{equationa}
\newcounter{remark}
\newcounter{example}
\newcounter{theorem}
\newcounter{proposition}
\newcounter{lemma}
\newcounter{corollary}
\newcounter{definition}
\def\theremark{\arabic{remark}}
\def\thetheorem{\arabic{theorem}}

\def\thedefinition{\arabic{definition}}

\newenvironment{eqalph}{\stepcounter{equation}
\setcounter{equationa}{\value{equation}} \setcounter{equation}{0}

\begin{eqnarray}}{\end{eqnarray}
\setcounter{equation}{\value{equationa}}}

\newcommand{\mar}[1]{}

\begin{document}
\hbox{}

{\parindent=0pt

{\large \bf Quantum mechanics with respect to different reference
frames}
\bigskip

{\sc L.Mangiarotti}

{\sl Department of Mathematics and Informatics, University of
Camerino, 62032 Camerino (MC), Italy}

\medskip

{\sc G.Sardanashvily}

{\sl Department of Theoretical Physics, Moscow State University,
117234 Moscow, Russia}

\bigskip
\bigskip

Geometric (Schr\"odinger) quantization of nonrelativistic
mechanics with respect to different reference frames is
considered. In classical nonrelativistic mechanics, a reference
frame is represented by a connection on a configuration space
fibered over a time axis $\Bbb R$. Under quantization, it yields a
connection on the quantum algebra of Schr\"odinger operators. The
operators of energy with respect to different reference frames are
examined.

 }

\bigskip
\bigskip

\noindent {\bf I. INTRODUCTION}
\bigskip

Since reference frame transformations are time-dependent, we are
in the case of time-dependent nonrelativistic mechanics. A problem
is that a Hamiltonian $\cH$ of time-dependent mechanics fails to
be a scalar function. Therefore, its quantization depends on a
reference frame. Quantized with respect to a given reference frame
$\G$, a Hamiltonian $\wh\cH$ coincides with the operator
$\wh\cE_\G$ of energy relative to this reference frame, but not to
other ones. Our goal is the relation (\ref{j21}) between operators
of energy with respect to different reference frames.

A configuration space of time-dependent mechanics is a smooth real
fiber bundle $\pi:Q\to \Bbb R$, where $Q$ is an oriented manifold
and $\Bbb R$ is a time axis. This configuration space is endowed
with bundle coordinates $(t,q^i)$ where $t$ is a Cartesian
coordinate on $\Bbb R$ possessing the transition functions
$t'=t+$const. Due to these transition functions, the base $\Bbb R$
is provided with the standard vector field $\dr_t$ and the
standard one-form $dt$. The velocity phase space of time-dependent
mechanics is the jet manifold $J^1Q$ of $Q\to \Bbb R$ coordinated
by $(t,q^i,q^i_t)$, and its momentum phase space is the vertical
cotangent bundle $V^*Q$ of $Q\to \Bbb R$ equipped with the
holonomic coordinates $(t,q^i,p_i=\dot q_i)$.$^{1-3}$

A fiber bundle over $Q\to\Bbb R$ is always trivial. Any
trivialization $Q\cong \Bbb R\times M$ yields a global reference
frame on $Q$ as follows. A reference frame in nonrelativistic
mechanics is defined as a connection
\mar{z270}\beq
\G=dt\ot (\dr_t +\G^i(t,q^j)\dr_i) \label{z270}
\eeq
on the configuration bundle $Q\to\Bbb R$.$^{1-5}$ Let
\mar{a1.10}\beq
\dr_\G = \dr_t + \G^i (t,q^j) \dr_i \label{a1.10}
\eeq
be the horizontal lift onto $Q$ of the standard vector field
$\dr_t$ on $\Bbb R$. One can think of its values as being
velocities of "observes" at points of $Q$. Accordingly, the
corresponding covariant differential
\mar{z279}\beq
D_\G: J^1Q\op\to_Q VQ, \qquad \dot q^i\circ D^\G =q^i_t-\G^i,
\label{z279}
\eeq
determines the relative velocities with respect to the reference
frame $\G$. These velocities are elements of the vertical tangent
bundle $VQ$ of $Q\to \Bbb R$ coordinated by $(t, q^i,\dot q_i)$. A
connection $\G$ (\ref{z270}) is called complete if the vector
field $\dr_\G$ (\ref{a1.10}) is complete.  Since  connections $\G$
on $Q\to\Bbb R$ are always flat, there is one-to-one
correspondence between them and the equivalence classes of atlases
of $Q\to\Bbb R$ with time-independent transition
functions.$^{1,5,6}$ Namely, $\G^i=0$ relative to bundle
coordinates of an associated atlas. These coordinates are called
$\G$-adapted. $\G$-adapted coordinates $(t,\ol q^j)$, a connection
$\G$ relative to arbitrary bundle coordinates $(t,q^i)$  reads
\mar{j10}\beq
\G^i(t,q^j)=\dr_t q^i(t,\ol q^j). \label{j10}
\eeq

In particular, every trivialization of a fiber bundle $Q\to \Bbb
R$ yields a complete connection, i.e.,  a reference frame on $Q$,
and {\it vice versa}. We call it a global reference frame and,
without a loss of generality, restrict our consideration to global
reference frames.

Let $T^*Q$ be the cotangent bundle of a configuration space $Q$
equipped with the holonomic coordinates $(t,q^i,p,p_i)$,
possessing the transition functions
\mar{z40}\beq
p'_i = \frac{\dr q^j}{\dr{q'}^i}p_j, \qquad p'=p+ \frac{\dr
q^j}{\dr t}p_j. \label{z40}
\eeq
This bundle plays the role of the homogeneous momentum phase space
of time-dependent mechanics.$^{3,7}$ It is equipped with the
canonical Liouville and symplectic forms
\mar{m91}\beq
\Xi=pdt +p_idq^i, \qquad \Om=d\Xi=dp\w dt +dp_i\w dq^i.
\label{m91}
\eeq
The corresponding Poisson bracket reads
\mar{m116}\beq
\{f,g\}_T =\dr^pf\dr_tg - \dr^pg\dr_tf +\dr^if\dr_ig-\dr^ig\dr_if,
\qquad f,g\in C^\infty(T^*Q). \label{m116}
\eeq

There is the trivial one-dimensional affine bundle
\mar{z11'}\beq
\zeta:T^*Q\to V^*Q. \label{z11'}
\eeq
It provides $V^*Q$ with the degenerate regular Poisson structure
\mar{m72}\beq
\{f,g\}_V = \dr^if\dr_ig-\dr^ig\dr_if, \qquad f,g\in
C^\infty(V^*Q), \label{m72}
\eeq
such that $\zeta$ is a Poisson morphism, i.e.,
\mar{m72'}\beq
\zeta^*\{f,g\}_V=\{\zeta^*f,\zeta^*g\}_T.\label{m72'}
\eeq
Due to this fact, we can quantize time-dependent mechanics in the
framework of geometric quantization of the Poisson manifold
$(V^*Q, \{,\}_V)$.$^{7-9}$

A Hamiltonian on the momentum phase space $V^*Q$ of time-dependent
mechanics is defined as a global section
\mar{ws513}\beq
\cH:V^*Q\to T^*Q, \qquad p=-\cH(t,q^j,p_j), \label{ws513}
\eeq
of the affine bundle $\zeta$ (\ref{z11'}). For instance, any
connection $\G$ (\ref{z270}) defines a Hamiltonian
\mar{j1}\beq
\cH_\G=\xi\circ\G: Q\to J^1Q\to T^*Q, \qquad \cH_\G=\G^ip_i,
\label{j1}
\eeq
due to the canonical imbedding $\xi:J^1Q\to T^*Q$. A glance at the
transformation law (\ref{z40}) shows that a Hamiltonian $\cH$
(\ref{ws513}) fails to be a function on $V^*Q$. Therefore, it can
not be quantized as an element of the Poisson algebra
$C^\infty(V^*Q)$ of smooth real functions on $V^*Q$. One can
overcome this difficulty if a trivialization $Q\cong\Bbb R\times
M$ (i.e., a reference frame) holds fixed,$^{10,11}$ but must
compare the results of independent quantizations with respect to
distinct reference frames.

In a different way, we have performed frame independent geometric
quantization of the Poisson algebra $C^\infty(T^*Q)$ on the
cotangent bundle $T^*X$.$^{7-9}$ It contains both the subalgebra
$\zeta^*C^\infty(V^*Q)$ and the function
\mar{mm16}\beq
\cH^*=\dr_t\rfloor(\Xi-\zeta^* h^*\Xi)=p+\cH, \label{mm16}
\eeq
which is a Hamiltonian, called the covariant Hamiltonian, on the
homogeneous momentum phase space $T^*Q$.$^{7,9}$ The monomorphism
$\zeta^*C^\infty(V^*Q)\to C^\infty(T^*Q)$ is prolonged to a
monomorphism of quantum algebras $\cA_V$ of $V^*Q$ and $\cA_T$ of
$T^*Q$. These algebras consist of affine functions of momenta on
$V^*Q$ and $T*Q$, respectively. They are represented by first
order differential operators (\ref{qq83}) and (\ref{qq82}) on the
$C^\infty(\Bbb R)$-module $E_R$ of complex half-forms on $Q$ whose
restriction to each fiber of $Q\to \Bbb R$ is of compact support.
Accordingly, the covariant Hamiltonian $\cH^*$, polynomial in
momenta, is quantized as an element of the universal enveloping
algebra of the Lie algebra $\cA_T$.

A problem is that the decomposition $\cH^*=p+\cH$ and the
corresponding splitting of quantum operators
\mar{j2}\beq
\wh\cH^*=\wh p+\wh\cH=-i\dr_t +\wh\cH \label{j2}
\eeq
are ill defined. At the same time, any reference frame $\G$ yields
the decomposition
\mar{j3}\beq
\cH^*=(p+\cH_\G) + (\cH-\cH_\G) = \cH^*_\G +\cE_\G, \label{j3}
\eeq
where $\cH_\G$ is the Hamiltonian (\ref{j1}), $\cH_\G^*$ is the
corresponding covariant Hamiltonian (\ref{j18}), and
\mar{j12}\beq
\cE_\G= \cH-\cH_\G=\cH-\G^ip_i \label{j12}
\eeq
is the energy relative to the reference frame $\G$.$^{1-3,5,12}$
Accordingly, we obtain the splitting
\mar{j4}\beq
\wh\cH^*=\wh\cH_\G^* +\wh\cE_\G, \label{j4}
\eeq
where $\wh\cE_\G$ is the operator of energy relative to the
reference frame $\G$ and $\wh\cH_\G^*$ (\ref{j19}) is a connection
(\ref{j17}) on the quantum algebra $\cA_V$ of $V^*Q$. The
splitting (\ref{j4}) results in a desired relation
\mar{j21}\beq
\wh\cH_\G^* +\wh\cE_\G=\wh\cH_{\G'}^* +\wh\cE_{\G'} \label{j21}
\eeq
between operators of energy with respect to different reference
frames.

Some examples are considered in Section V.

\bigskip
\bigskip

\noindent {\bf II. CLASSICAL EVOLUTION OPERATOR AND ENERGY
FUNCTIONS}
\bigskip

In contrast with autonomous mechanics, the Poisson structure
(\ref{m72}) fails to provide any dynamic equation on the momentum
phase space $V^*Q$ because Hamiltonian vector fields
\mar{m73}\beq
\vt_f = \dr^if\dr_i- \dr_if\dr^i, \qquad f\in C^\infty(V^*Q),
\label{m73}
\eeq
of functions on $V^*Q$ are vertical. Moreover, a Hamiltonian $\cH$
of time-dependent mechanics is not a function on $V^*Q$ and,
therefore, its Hamiltonian vector field with respect to the
Poisson structure  (\ref{m72}) is not defined. Dynamics of
time-dependent mechanics is described as follows.$^{1-3}$

A Hamiltonian $\cH$ (\ref{ws513}) definess the pull-back
Hamiltonian form
\mar{b4210}\beq
H=\cH^*\Xi= p_k dq^k -\cH dt  \label{b4210}
\eeq
on $V^*Q$, which is the well-known invariant of
Poincar\'e--Cartan. Given a Hamiltonian form $H$ (\ref{b4210}),
there exists a unique horizontal vector field $\g_H$ (i.e.,
$\g_H\rfloor dt=1$) on $V^*Q$ such that $\g_H\rfloor dH=0$. It
reads
\mar{z3}\beq
\g_H=\dr_t + \dr^k\cH\dr_k- \dr_k\cH\dr^k. \label{z3}
\eeq
This vector field, called the Hamilton vector field, yields the
first order Hamilton equations
\mar{z20}\beq
q^k_t-\dr^k\cH=0, \qquad  p_{tk}+\dr_k\cH=0 \label{z20}
\eeq
on $V^*Q\to\Bbb R$, where $(t,q^k,p_k,q^k_t,p_{tk})$ are
coordinates on the first order jet manifold $J^1V^*Q$ of
$V^*Q\to\Bbb R$. Accordingly, the Lie derivative
\mar{ws516}\beq
\bL_{\g_H} f=\g_H\rfloor df=(\dr_t + \dr^k\cH\dr_k-
\dr_k\cH\dr^k)f \label{ws516}
\eeq
of functions $f\in C^\infty(V^*Q)$ along the Hamilton vector field
$\g_H$ (\ref{z3}) is called the evolution operator. It takes the
form
\be
\bL_{\g_H} f\ap d_tf=(\dr_t +q^i_t\dr_i +p_{ti}\dr^i)f
\ee
on-shell (i.e., on the submanifold of $J^1V^*Q$ defined by the
Hamilton equations (\ref{z20})). If $\bL_{\g_H}=0$, then $d_tf\ap
0$, i.e., a function $f$ is an integral of motion. The evolution
operator takes the local form
\be
\bL_{\g_H} f=\dr_tf +\{\cH,f\}_V,
\ee
but the Poisson bracket $\{\cH,f\}_V$ is ill defined because it is
not a function under coordinate transformations.

In order to overcome this difficulty and quantize the evolution
operator, we use the fact that a time-dependent Hamiltonian system
on the momentum phase space $V^*Q$ can be extended to an
autonomous Hamiltonian system on the homogeneous momentum phase
space $T^*Q$ with the covariant Hamiltonian $\cH^*$
(\ref{mm16}).$^{3,13,14}$ The Hamiltonian vector field of $\cH^*$
with respect to the Poisson bracket (\ref{m116}) on $T^*Q$ reads
\mar{z5}\beq
\g_T=\dr_t -\dr_t\cH\dr^p+ \dr^k\cH\dr_k- \dr_k\cH\dr^k.
\label{z5}
\eeq
It is projected onto the Hamilton vector field $\g_H$ (\ref{z3})
on $V^*Q$ so that the relation
\mar{ws525}\beq
\zeta^*(\bL_{\g_H}f)=\{\cH^*,\zeta^*f\}_T, \qquad f\in
C^\infty(V^*Q), \label{ws525}
\eeq
holds. In particular, a function $f\in C^\infty(V^*Q)$ is an
integral of motion iff the bracket (\ref{ws525}) vanishes.

Due to the relation (\ref{ws525}), we can quantize the pull-back
of the evolution operator (\ref{ws516}) onto $T^*Q$ in the
framework of geometric quantization of the symplectic manifold
$(T^*Q,\Om)$.

It is readily observed that the pull-back of a Hamiltonian form
$H$ (\ref{b4210}) onto the jet manifold $J^1V^*Q$ is the
Poincar\'e--Cartan form of the Lagrangian
\mar{Q33}\beq
L_H=h_0(H) = (p_iq^i_t - \cH)dt, \qquad h_0(dq^i)=q^i_tdt,
\label{Q33}
\eeq
on $J^1V^*Q$, and that the Hamilton equations (\ref{z20}) are
exactly the Lagrange equations of this Lagrangian.$^1$ Using this
fact, one can obtain conservation laws in time-dependent
Hamiltonian mechanics as follows.$^{1,2,5}$

Any vector field $u=u^t\dr_t+u^i\dr_i$ on a configuration space
$Q$ gives rise to the vector field
\mar{gm513}\beq
\wt u=u^t\dr_t + u^i\dr_i - \dr_i u^j p_j\dr^i \label{gm513}
\eeq
on the momentum phase space $V^*Q$. The equality
\mar{mm24}\beq
\bL_{\wt u}H= \bL_{J^1\wt u}L_H \label{mm24}
\eeq
holds. Then the first variational formula applied to the
Lagrangian $L_H$ (\ref{Q33}) leads to the weak identity
\be
\bL_{\wt u}H\ap d_t(u\rfloor H)dt
\ee
on-shell. If the Lie derivative (\ref{mm24}) vanishes, we come to
the weak conservation law $0\ap d_t(u\rfloor H)dt$ of the function
\mar{b4306}\beq
\gT_u= u\rfloor dH= p_iu^i - u^t\cH. \label{b4306}
\eeq
By analogy with field theory, this function is called the symmetry
current along $u$.  Any conserved symmetry current $\gT_u$
(\ref{b4306}) is an integral of motion.

For instance, put $u=-\dr_\G$ (\ref{a1.10}). The corresponding
symmetry current $\gT_{-\dr_\G}$ (\ref{b4306}) is the energy
function $\cE_\G$ (\ref{j12}) relative to a reference frame
$\G$.$^{1,5,15}$ A glance at the expression (\ref{j12}) shows that
$\cE_\G=\cH$ with respect to $\G$-adapted coordinates.

Note that, given a reference frame $\G$, one should solve the
equations
\mar{gm300}\beq
\G^i(t, q^j(t,\ol q^a))=\frac{\dr q^i(t,\ol q^a)}{\dr t}, \qquad
 \frac{\dr \ol q^a(t,q^j)}{\dr q^i}\G^i(t,q^j) +\frac{\dr \ol
q^a(t,q^j)}{\dr t}=0 \label{gm300}
\eeq
in order to find the coordinates $(t,\ol q^a)$ adapted to
$\G$.$^{1,5}$ In particular, one can show that components $\G^i$
of $\G$ are affine in coordinates $q^i$ iff transition functions
between coordinates $q^i$ and $\ol q^a$ are affine, i.e.,
$q^i=a^i_k(t)\ol q^k +b^i(t)$.$^1$ Examples in Section V
illustrate this fact.

With the energy function $\cE_\G$, we obtain the decomposition
(\ref{j3}) of a covariant Hamiltonian. It follows that
\mar{j34}\beq
\zeta^*(d_t\cE_\G)\ap
\{\cH^*,\zeta^*\cE_\G\}_T=\zeta^*(\dr_t\cE_\G). \label{j34}
\eeq
A Hamiltonian system is called conservative if there exists a
reference frame $\G$ such that, written with respect to
$\G$-adapted coordinates, its Hamiltonian $\cH=\cE_\G$ is
time-independent. A glance at the equality (\ref{j34}) shows that,
in this case, $\cE_\G$ is an integral of motion, and {\it vice
versa}.

Given different reference frames $\G$ and $\G'$, the decomposition
(\ref{j3}) leads at once to the relation
\mar{j45}\beq
\cE_{\G'}=\cE_\G + \cH_\G -\cH_{\G'}=\cE_\G + (\G^i -\G'^i)p_i.
\label{j45}
\eeq

\bigskip
\bigskip

\noindent {\bf III. QUANTUM TIME-DEPENDENT MECHANICS}
\bigskip

Following Refs. [7-9], we quantize time-dependent mechanics in the
framework of geometric quantization of the cotangent bundle $T^*Q$
with respect to its vertical polarization given by the vertical
tangent bundle $VT^*Q$ of $T^*Q\to Q$. Note that polarization of
$T^*Q$ need not induce polarization of $V^*Q$, unless it contains
the vertical cotangent bundle $V_\zeta T^*Q$ of the fiber bundle
$\zeta$ (\ref{z11'}) spanned by vectors $\dr_p$. The above
mentioned vertical polarization is unique canonical real
polarization of $T^*Q$ satisfying this condition. With this
polarization, the monomorphism of Poisson algebras
\be
\zeta^*: (C^\infty(V^*Q),\{,\}_V) \to (C^\infty(T^*Q),\{,\}_T)
\ee
defined by the relation (\ref{m72'}) is prolonged to a
monomorphism $\cA_V\to\cA_T$ of quantum algebras $\cA_V$ of $V^*Q$
and $\cA_T$ of $T^*Q$. A result is the following (see Appendix for
the quantization technique).

For the sake of simplicity, we assume that $Q$ is a simply
connected manifold and that the cohomology group $H^2(Q,\Bbb Z_2)$
is trivial.$^9$ Otherwise, there are nonequivalent quantizations.
For the sake of convenience, the compact notation $(q^\la,
p_\la)$, $q^0=t$, $p_0=p$, is further used.

Quantum algebras $\cA_V$ and $\cA_T$ consist of affine functions
of momenta on the cotangent bundle $V^*Q$ and the vertical
cotangent bundle $T^*Q$, respectively. We obtain their
Schr\"odinger representations by first order linear differential
operators
\mar{qq82,83}\ben
&& \wh f\rho=(-i\bL_{a^k\dr_k} +b)\rho= (-ia^k\dr_k
-\frac{i}{2}\dr_k a^k+ b) \rho, \qquad f=a^k(q^\mu)p_k +
b(q^\mu)\in \cA_V, \label{qq83}\\
&& \wh f\rho=(-i\bL_{a^\la\dr_\la} +b)\rho= (-ia^\la\dr_\la
-\frac{i}{2}\dr_\la a^\la+ b) \rho, \qquad f=a^\la(q^\mu)p_\la +
b(q^\mu)\in \cA_T, \label{qq82}
\een
which act in the $C^\infty(\Bbb R)$-module $E_R$ of complex
half-forms $\rho$ on $Q$ whose restriction to each fiber $Q_t$ of
$Q\to \Bbb R$ is of compact support. A glance at the expressions
(\ref{qq83}) and (\ref{qq82}) shows that $\wh f$ (\ref{qq83}) is
the representation of $\cA_V$ as a subalgebra of $\cA_T$.

It is readily justified that the operators (\ref{qq83}) and
(\ref{qq82}) satisfy the Dirac condition
\mar{j20}\beq
[\wh f,\wh f']=-i\wh{\{f,f'\}_T} \label{j20}
\eeq
and, thus, form real Lie algebras. Moreover, the quantum operators
of functions $f(t)\in C^\infty(\Bbb R)$ depending only on time are
reduced to multiplications $\wh f\rho=f\rho$, and they commute
with any element (\ref{qq83}) of the quantum algebra $\cA_V$. It
follows that $\cA_V$ is a Lie $C^\infty(\Bbb R)$-algebra. Its
representation (\ref{qq83}) defines the instantwise quantization
of $\cA_V$ as follows.

Geometric quantization of the Poisson manifold $(V^*Q,\{,\}_V)$
yields geometric quantization of its symplectic leaves which are
fibers $i_t:V^*_tQ\to V^*Q$, $t\in\Bbb R$ of the fiber bundle
$V^*Q\to\Bbb R$ provided with the symplectic structure
\mar{kk}\beq
\Om_t=(\cH\circ i_t)^*\Om=dp_k\w dq^k, \label{kk}
\eeq
where $\cH$ is an arbitrary section of the fiber bundle $\zeta$
(\ref{z11'}). The associated quantum algebra $\cA_t$ consists of
functions  on $V^*_tQ$ which are affine in momenta. It is
represented by Hermitian operators
\mar{qq83'}\beq
\wh f_t\rho_t=(-i\bL_{a^k\dr_k} +b)\rho_t= (-ia^k\dr_k
-\frac{i}{2}\dr_k a^k+ b) \rho_t, \qquad f_t=a^k(q^i)p_k +
b(q^i)\in \cA_t, \label{qq83'}
\eeq
in the pre-Hilbert space $E_t$ of half-forms of compact support on
$Q_t$ equipped with the Hermitian metric
\mar{j29}\beq
\lng \rho'_t|\rho_t\rng_t=\op\int_{Q_t}\ol\rho'_t \rho_t.
\label{j29}
\eeq
Spaces $E_t$, $t\in \Bbb R$, are assembled into a trivial
pre-Hilbert bundle over $\Bbb R$. One can show that the
restriction to $V^*_tQ$ of any element of the quantum algebra
$\cA_V$ belongs to $\cA_t$ and, conversely, any element of $\cA_t$
is of this type.  Thus, $\cA_t=i_t^*\cA_V$. Any half-form $\rho\in
E_R$ on $Q$ yields a half-form of compact support on $Q_t$. Given
an element $f\in \cA_V$ and its pull-back $f_t=i_t^*f\in \cA_t$,
we obtain from the formulas (\ref{qq83}) and (\ref{qq83'}) that
$\wh f\rho\circ i_t=\wh f_t(\rho\circ i_t)$.

As was mentioned above, the $C^\infty(\Bbb R)$-module $E_R$ is
also the carrier space for the quantum algebra $\cA_T$, but its
action on $E_R$ is not instantwise.

\bigskip
\bigskip

\noindent {\bf IV. ENERGY OPERATORS}
\bigskip

Turn now to quantization of the covariant Hamiltonian $\cH^*$
(\ref{mm16}). A problem is that, in the framework of the
Schr\"odinger quantization, it fails to belong to the quantum
algebra $\cA_T$, unless it is affine in momenta. Let us restrict
our consideration to the physically relevant case of $\cH^*$
polynomial in momenta. One can show that, in this case, $\cH^*$ is
decomposed in a finite sum of products of elements of the algebra
$\cA_T$, though this decomposition by no means is unique.$^{7,9}$
Provided with such a decomposition, $\cH^*$ is an element of the
universal enveloping algebra $\ol\cA_T$ of the Lie algebra
$\cA_T$. To be more precise, it belongs to $\cA_T+\ol\cA_V$, where
$\ol\cA_V$ is the universal enveloping algebra of the Lie algebra
$\cA_V\subset \cA_T$. Since the Dirac condition (\ref{j20}) holds,
the Schr\"odinger representation of the Lie algebras $\cA_T$ and
$\cA_V$ is naturally extended to their enveloping algebras
$\ol\cA_T$ and $\ol\cA_V$, and provides the quantization
$\wh\cH^*$ of a covariant Hamiltonian $\cH^*$.

For instance, let $\cH_\G$ (\ref{j1}) be a Hamiltonian defined by
a connection $\G$ on $V^*Q$. The corresponding covariant
Hamiltonian
\mar{j18}\beq
\cH^*_\G=p+\cH_\G=p +\G^kp_k \label{j18}
\eeq
belongs to the quantum algebra $\cA_T$. Its quantization
(\ref{qq83}) reads
\mar{j19}\beq
\wh\cH^*_\G =-i\dr_t -i\G^k\dr_k -\frac{i}{2}\dr_k\G^k.
\label{j19}
\eeq
Written with respect to $\G$-adapted coordinates, it takes the
form $\wh\cH^*_\G =-i\dr_t$.

Given an operator $\wh\cH^*$, the bracket
\mar{qq120}\beq
\wh\nabla\wh f= i[\wh \cH^*,\wh f] \label{qq120}
\eeq
defines a derivation of the algebra $\ol\cA_V$. Moreover, since
$\wh p=-i\dr_t$, the derivation (\ref{qq120}) obeys the Leibniz
rule
\be
\wh\nabla (g\wh f)=\dr_t g\wh f + g\nabla \wh f, \qquad g\in
C^\infty(\Bbb R).
\ee
Therefore, it is a connection on the $C^\infty(\Bbb R)$-module
$\ol\cA_V$.$^{5,9}$ One says that $\wh f$ is parallel with respect
to the connection (\ref{qq120}) if
\mar{qq151}\beq
[\wh \cH^*,\wh f]=0. \label{qq151}
\eeq
One can think of this equality as being the Heisenberg equation in
time-dependent mechanics, while the quantum constraint
\mar{gg}\beq
\wh\cH^*\rho=0, \qquad \rho\in E_R, \label{gg}
\eeq
plays a role of the Schr\"odinger equation.

For instance, any connection (i.e., a reference frame) $\G$
(\ref{a1.10}) on a configuration bundle $Q\to\Bbb R$ induces the
connection
\mar{j17}\beq
\wh\nabla_\G\wh f=i[\wh\cH_\G,\wh f] \label{j17}
\eeq
on the algebra $\ol\cA_V$ which is also a connection on the
quantum algebra $\cA_V\subset \ol\cA_V$. The corresponding
Schr\"odinger equation (\ref{gg}) reads
\be
-i(\dr_t +\G^k\dr_k +\frac12\dr_k\G^k)\rho=0.
\ee
Its solutions are half-forms $\rho\in E_R$ which, written relative
to $\G$-adapted coordinates $(t,\ol q^j)$, are time-independent,
i.e., $\rho=\rho(\ol q^j)$.

Given a reference frame $\G$, the energy function $\cE_\G$ is
quantized as $\wh \cE_\G=\wh\cH^*- \wh\cH^*_\G$. One can think of
$\wh \cE_\G$ as being an operator of energy with respect to the
reference frame $\G$. As a consequence, the Schr\"odinger equation
(\ref{gg}) reads
\mar{j33}\beq
(\wh\cH_\G +\wh\cE_\G)\rho=-i(\dr_t +\G^k\dr_k
+\frac12\dr_k\G^k)\rho +\wh\cE_\G\rho=0. \label{j33}
\eeq
Let a Hamiltonian system be conservative and $\G$ a reference
frame such that the energy function $\cE_\G$ is time-independent
relative to $\G$-adapted coordinates. In this case, the
Schr\"odinger equation (\ref{j33}) takes the familiar form
\mar{j35}\beq
(-i\dr_t +\wh\cE_\G)\rho=0. \label{j35}
\eeq
It follows from the Heisenberg equation (\ref{qq151}) that a
Hamiltonian system is conservative iff there exists a reference
frame $\G$ such that
\be
[\wh\cH^*,\wh\cE_\G]=0.
\ee

Given different reference frames, operators of energy $\wh\cE_\G$
and $\wh\cE_{\G'}$ obey the relation (\ref{j21}) taking the form
\mar{j46}\beq
\wh\cE_{\G'}=\wh\cE_\G -i(\G^k-\G'^k)\dr_k
-\frac{i}{2}\dr_k(\G^k-\G'^k). \label{j46}
\eeq
In particular, let $\wh\cE_\G$ be a time-independent energy
operator of a conservative system, and let $\rho_E$ be its
eigenstate of eigenvalue $E$, i.e., $\wh\cE_\G\rho_E=E\rho_E$.
Then the energy of this state relative to a reference frame $\G'$
at an instant $t$ is
\be
&& \lng \rho_E|\wh\cE_{\G'}\rho_E\rng =   E+i\lng
\rho_E|(\G'^k\dr_k +\frac12\dr_k\G'^k)\rho_E\rng_t= \\
&& \qquad  E+ i\op\int_{Q_t} \ol\rho_E(\G'^k(q^j,t)\dr_k
+\frac12\dr_k\G'^k(q^j,t))\rho_E.
\ee

\bigskip
\bigskip

\noindent {\bf V. EXAMPLES}
\bigskip

Let us consider a Hamiltonian system on $Q=\Bbb R\times U$, where
$U\subset\Bbb R^m$ is an open domain equipped with coordinates
$(q^i)$. These coordinates yield a reference frame on $Q$ given by
the connection $\G$ such that $\G^i=0$ with respect to these
coordinates. Let it be a conservative Hamiltonian system whose
energy function $\cE_\G$, written relative to coordinates
$(t,q^i)$, is time-independent. Let us consider a different
reference frame on $Q$ given by the connection
\mar{j40}\beq
\G'=dt\ot(\dr_t+ G^i\dr_i), \qquad G^i={\rm const}, \label{j40}
\eeq
on $Q$. The $\G'$-adapted coordinates $(t,q'^j)$ obey the
equations (\ref{gm300}) which read
\mar{j41}\beq
G^i=\frac{\dr q^i(t, q'^j)}{\dr t}, \qquad \frac{\dr
q'^j(t,q^i)}{\dr q^k}G^k +\frac{\dr  q'^j(t,q^i)}{\dr t}=0.
\label{j41}
\eeq
We obtain $q'^i=q^i -G^it$. For instance, this is the case of
inertial frames.$^1$ Given by the relation (\ref{j45}), the energy
function relative to the reference frame $\G'$ (\ref{j40}) reads
\be
\cE_{\G'}=\cE_\G-G^k p_k.
\ee
Accordingly, the relation (\ref{j46}) between operators of energy
$\wh \cE_{\G'}$ and $\wh\cE_\G$ takes the form
\mar{j48}\beq
\wh \cE_{\G'}=\wh\cE_\G +iG^k\dr_k. \label{j48}
\eeq
Let $\rho_E$ be an eigenstate of the energy operator $\wh\cE_\G$.
Then its energy with respect to the reference frame $\G'$
(\ref{j40}) is $E-G^kP_k$, where $P_k=\lng\rho_E|\wh
p_k\rho_E\rng_t$ are momenta of this state. This energy is
time-independent.

In particular, the following condition holds in many physical
models. Given an eigenstate $\rho_E$ of the energy operator
$\wh\cE_\G$ and a reference frame $\G'$ (\ref{j40}), there is the
equality
\be
\wh \cE_{\G'}(\wh p^j, q^j)\rho_E=(\wh\cE_\G(\wh p^j, q^j) -
G^k(q^j)\wh p_k)\rho_E= (\wh\cE_\G(\wh p^j+A_j, q^j) + B)\rho_E,
\qquad A_j,B={\rm const}.
\ee
Then $\exp(-iA_jq^j)\rho_E$ is an eigenstate of the energy
operator $\wh \cE_{\G'}$ possessing the eigenvalue $E+ B$.

For instance, any Hamiltonian
\be
\wh\cH=\wh\cE_\G=\frac12(m^{-1})^{ij}\wh p_i\circ\wh p_j + V(q^j)
\ee
quadratic in momenta $\wh p_i$ with a nondegenerate constant mass
tensor $m^{ij}$ obeys this condition.$^1$ Namely, we have
\be
A_i=-m_{ij}G^j, \qquad B=-\frac12 m_{ij}G^iG^j.
\ee

Let us consider a massive point particle in an Euclidean space
$\Bbb R^3$ in the presence of a central potential $V(r)$. Let
$\Bbb R^3$ ,be equipped with the spherical coordinates
$(r,\f,\theta)$. These coordinates define an inertial reference
frame $\G$ such that $\G^r=\G^\f=\G^\theta=0$. The Hamiltonian of
the above mentioned particle with respect to this reference frame
reads
\mar{j50}\beq
\wh\cH=\wh\cE_\G=\frac{1}{m}(-\frac{1}{r}\dr_r -\frac12\dr^2_r +
\frac{\wh I^2}{r^2}) +V(r), \label{j50}
\eeq
where $\wh I$ is the square of the angular momentum operator. Let
us consider a rotatory reference frame $\G'^\f=\om$, $\om=$const,
given by the adapted coordinates $(r,\f'=\f -\om t,\theta)$. The
operator of energy relative to this reference frame is
\mar{j51}\beq
\wh\cE_{\G'}=\wh\cE_\G + i\om\dr_\f. \label{j51}
\eeq
Let $\rho_{E,n,l}$ be an eigenstate of the energy operator
$\wh\cE_\G$ (\ref{j50}) posessing its eigenvalue $E$, the
eigenvalue $n$ of the angular momentum operator $\wh I_3=\wh
p_\f$, and the eigenvalue $l(l+1)$ of the operator $\wh I^2$. Then
$\rho_{E,n,l}$ is also an eigenstate of the energy operator
$\wh\cE_{\G'}$ with the eigenvalue $E'=E -n\om$.

\bigskip
\bigskip

\noindent {\bf VI. APPENDIX}
\bigskip

We start with the standard prequantization of the cotangent bundle
$T^*Q$.  Since the symplectic form $\Om$ (\ref{m91}) on $T^*Q$ is
exact, a prequantization bundle is the trivial complex line bundle
\mar{qm501}\beq
C=T^*Q\times\Bbb C\to T^*Q,  \label{qm501}
\eeq
coordinated by $(q^\la,p_\la,c)$. It is provided with the
admissible linear connection
\mar{qm502}\beq
A=dp_\la\ot\dr^\la +dq^\la\ot(\dr_\la+ ip_\la c\dr_c)
\label{qm502}
\eeq
whose curvature equals $-i\Om$. The $A$-invariant Hermitian fiber
metric on $C$ is $g(c,c')=c\ol c'$. The covariant derivative of
sections $s$ of the prequantization bundle $C$ (\ref{qm501})
relative to the connection $A$ (\ref{qm502}) along a vector field
$u$ on $T^*Q$ reads
\be
\nabla_u s= u^\la(\dr_\la - i p_\la)s +u_\la\dr^\la s.
\ee
Given a function $f\in C^\infty(T^*Q)$ and its Hamiltonian vector
field $u_f=\dr^\la f\dr_\la -\dr_\la f\dr^\la$, one assigns to $f$
the first order differential operator
\mar{qm504}\beq
\wh f(s) =-i(\nabla_{u_f} + if)s=[-i(\dr^\la f\dr_\la -\dr_\la
f\dr^\la) + (f -p_\la\dr^\la f)]s \label{qm504}
\eeq
on sections $s$ of the prequantization bundle $C$ (\ref{qm501}).
These operators obey the Dirac condition (\ref{j20}) and, thus,
form a prequantum real Lie algebra $\wh C^\infty(T^*Q)$ of $T^*Q$.

Let us turn to prequantization of the Poisson manifold
$(V^*Q,\{,\}_V)$. Its Poisson bivector field $w=\dr^k\w\dr_k$ is
exact. Therefore, a prequantization bundle is the trivial complex
line bundle
\mar{qq43}\beq
C_V=V^*Q\times \Bbb C\to V^*Q. \label{qq43}
\eeq
Since the bundles $C$ (\ref{qm501}) and $C_V$ (\ref{qq43}) are
trivial, $C$ can be seen as the pull-back $\zeta^*C_V$ of $C_V$,
while $C_V$ is isomorphic to the pull-back $\cH^*C$ of $C$ with
respect to some section $\cH$ (\ref{ws513}) of the affine bundle
(\ref{z11'}). Because the covariant derivative of the connection
$A$ (\ref{qm502}) along the fibers of $\zeta$ (\ref{z11'}) is
trivial, let us consider the pull-back
\mar{qq44}\beq
\cH^*A=dp_k\ot\dr^k +dq^k\ot(\dr_k+ ip_k c\dr_c)+ dt\ot(\dr_t -
i\cH c\dr_c) \label{qq44}
\eeq
of $A$ onto $C_V\to V^*Q$. This connection defines the
contravariant derivative
\mar{qq45}\beq
\nabla_\f s_V = \nabla_{w^\sharp\f}s_V, \qquad
w^\sharp\f=\f^k\dr_k-\f_k\dr^k, \label{qq45}
\eeq
of sections $s_V$ of $C_V\to V^*Q$ along one-forms $\f$ on $V^*Q$.
The contravariant derivative (\ref{qq45}) corresponds to a
contravariant connection $A_V$ on the line bundle $C_V\to
V^*Q$.$^{16}$ Since vector fields $w^\sharp\f$ are vertical on
$V^*Q\to \Bbb R$, this contravariant connection does not depend on
a choice of $\cH$. By virtue of the relation (\ref{qq45}), the
curvature bivector of $A_V$ equals $-iw$, i.e., $A_V$ is an
admissible connection on $V^*Q$. Then the Kostant--Souriau formula
\mar{qq46}\beq
\wh f(s_V) =(-i\nabla_{\vt_f}+f)s_V=[-i(\dr^kf\dr_k-\dr_kf\dr^k)+
(f -p_k\dr^k f)]s_V, \label{qq46}
\eeq
defines prequantization of the Poisson manifold $V^*Q$. In
particular, the prequantum operators of functions $f(t)\in
C^\infty(\Bbb R)$ of time are reduced to multiplications $\wh
f(s_V)=fs_V$,  and they commute with any element (\ref{qq46}) of
the quantum algebra $\cA_V$. Consequently, the prequantum algebra
$\wh C^\infty(V^*Q)$ of $V^*Q$ is a Lie $C^\infty(\Bbb
R)$-algebra. It is readily observed that the prequantum operator
$\wh f$ (\ref{qq46}) coincides with the prequantum operator
$\wh{\zeta^*f}$ (\ref{qm504}) restricted to the pull-back sections
$s=\zeta^*s_V$. Thus, the above mentioned prequantization of the
Poisson algebra $C^\infty(V^*Q)$ is equivalent to its
prequantization as a subalgebra of the Poisson algebra
$C^\infty(T^*Q)$. Since the line bundles $C$ (\ref{qm501}) and
$C_V$ (\ref{qq43}) are trivial, their sections are smooth complex
functions on $T^*Q$ and $V^*Q$, respectively. Then the prequantum
operators (\ref{qm504}) and (\ref{qq46}) can be written in the
form
\mar{jj}\beq
\wh f=-i\bL_{u_f} +(f- \bL_\vt f), \label{jj}
\eeq
where $\vt$ is either the Liouville vector field
$\vt=p_\la\dr^\la$ on $T^*Q$  or $\vt=p_k\dr^k$ on $V^*Q$.

Given compatible prequantizations of $T^*Q$ and $V^*Q$, let us
construct their compatible polarizations and quantizations. Recall
that polarization of a Poisson manifold $(Z,\{,\})$ is defined as
a sheaf $\bT^*$ of germs of complex functions on $Z$ whose stalks
$\bT^*_z$, $z\in Z$, are Abelian algebras with respect to the
Poisson bracket $\{,\}$.$^{16,17}$. A quantum algebra $\cA$
associated to the Poisson polarization $\bT^*$ is a subalgebra of
the Poisson algebra $C^\infty(Z)$ which consists of functions $f$
such that $\{f,\bT^*(Z)\}\subset \bT^*(Z)$. Polarization of a
symplectic manifold yields its Poisson one.

Let $\bT^*$ be a polarization of the Poisson manifold
$(T^*Q,\{,\}_T)$. Its direct image in $V^*Q$ with respect to the
fibration $\zeta$ (\ref{z11'}) is polarization of the Poisson
manifold $(V^*Q,\{,\}_V)$ if elements of $\bT^*$ are germs of
functions constant on fibers of $\zeta$, i.e., independent of the
momentum coordinate $p$. It follows that the corresponding
symplectic polarization $\bT$ of $T^*Q$ is vertical with respect
to $\zeta$. The polarization $\bT=VT^*Q$ of $T^*Q$ obeys this
condition. The associated quantum algebra $\cA_T\subset
C^\infty(T^*Q)$ consists of functions which are affine in momenta
$p_\la$. It acts by operators (\ref{jj}) on the space of smooth
complex functions $s$ on $T^*Q$ which fulfill the relation
$\nabla_u s=0$ for any vertical vector field $u$ on $T^*Q\to Q$.
Clearly, these functions are the pull-back of complex functions on
$Q$ along the fibration $T^*Q\to Q$. Following the metaplectic
correction procedure, we come to complex half-forms on $Q$ which
are sections of the complex line bundle $\cD_{1/2}\to Q$ with the
transition functions $c'=Sc$ such that $S\ol S$ is the Jacobian of
coordinate transition functions on $Q$. Then the formula
(\ref{jj}), where $\bL$ is the Lie derivative of half-forms,
defines the Schr\"odinger representation of the real Lie algebra
$\cA_T$ by operators (\ref{qq82}) in the space $\cD_{1/2}(Q)$ of
complex half-forms $\rho$ on $Q$. These operators are Hermitian on
the pre-Hilbert space $E\subset \cD_{1/2}(Q)$ of half-forms of
compact support provided with the nondegenerate Hermitian form
\mar{nn}\beq
\lng \rho'|\rho\rng= \op\int_Q \ol\rho'\rho. \label{nn}
\eeq

The vertical polarization of $T^*Q$ induces the polarization
$\bT_V^*$ of the Poisson manifold $V^*Q$ which contains the germs
of functions constant on fibers of $V^*Q\to Q$. The associated
quantum algebra $\cA_V$ consists of functions on $V^*Q$ which are
affine in momenta. It acts by operators (\ref{jj}) on the space of
smooth complex functions $s_V$ on $V^*Q$ which fulfill the
relation $\nabla_u s_V=0$ for any vertical vector field $u$ on
$V^*Q\to Q$. These functions are the pull-back of complex
functions on $Q$ along the fibration $V^*Q\to Q$. Similarly to the
case of $\cA_T$, we obtain the Schr\"odinger representation of
$\cA_V$ by the operators (\ref{qq83}) on half-forms on $Q$.
Moreover, a glance at the expressions (\ref{qq83}) and
(\ref{qq82}) shows that (\ref{qq83}) is the representation of
$\cA_V$ as a subalgebra of $\cA_T$.

However, the physical relevance of the pre-Hilbert space $E$ with
the Hermitian form (\ref{nn}) is open to question. This Hermitian
form implies integration over time, though the time plays a role
of the classical evolution parameter in quantum mechanics. Since
the representation (\ref{qq83}) preserves the structure of $\cA_V$
as a Lie $C^\infty(\Bbb R)$-algebra, let us show that this
representation defines the instantwise quantization of $\cA_V$.

The prequantization (\ref{qq46}) of the Poisson manifold $V^*Q$
yields prequantization of its symplectic leaf $i_t:V_t^*Q\to
V^*Q$, $t\in\Bbb R$, endowed with the symplectic form (\ref{kk}).
Since $w^\sharp\f$ is a vertical vector field on $V^*Q\to\Bbb R$
for any one-form $\f$ on $V^*Q$, the contravariant derivative
(\ref{qq45}) defines a connection along each fiber $V_t^*Q$,
$t\in\Bbb R$, of the Poisson bundle $V^*Q\to\Bbb R$. It is the
pull-back
\be
A_t=i^*_t h^*A= dp_k\ot\dr^k +dq^k\ot(\dr_k+ ip_k c\dr_c)
\ee
of the connection $\cH^*A$ (\ref{qq44}) onto the trivial pull-back
line bundle
\be
i_t^*C_V=V_t^*Q\times\Bbb C\to V_t^*Q.
\ee
It is readily observed that this connection is admissible for the
symplectic structure (\ref{kk}) on $V^*_tQ$, and provides
prequantization of the symplectic manifold $(V_t^*Q,\Om_t)$ by the
formula
\mar{vv}\beq
\wh f_t
=-i\bL_{\vt_{f_t}}+(f_t-\bL_{\vt_t})=-i(\dr^kf_t\dr_k-\dr_kf_t\dr^k)+
(f_t -p_k\dr^k f_t),  \label{vv}
\eeq
where $\vt_{f_t}= \dr^kf_t\dr_k-\dr_kf_t\dr^k$ is the Hamiltonian
vector field of a function $f_t$ on $V^*_tQ$ with respect to the
symplectic form $\Om_t$ (\ref{kk}). The operators (\ref{vv}) act
on smooth complex functions $s_t$ on $V^*_tQ$. In particular, let
$f_t$, $s_t$ and $(\wh fs)_t$ be the restriction to $V^*_tQ$ of a
real function $f$ and complex functions $s$ and $\wh f(s)$ on
$V^*Q$, respectively. We obtain from the formulas (\ref{qq46}) and
(\ref{vv}) that $(\wh f s)_t=\wh f_t s_t$. It follows that the
prequantization (\ref{qq46}) of the Poisson manifold $V^*Q$ is a
fiberwise prequantization.

Let $\bT_V^*$ be the above mentioned polarization of the Poisson
manifold $V^*Q$. It yields the pull-back polarization
$\bT_t^*=i^*_t\bT^*_V$ of a fiber $V^*_tQ$ with respect to the
Poisson morphism $i_t:V^*_tQ\to V^*Q$. The corresponding
distribution $\bT_t$ coincides with the vertical tangent bundle of
the fiber bundle $V^*_tQ\to Q_t$. The associated quantum algebra
$\cA_t$ consists of functions  on $V^*Q_t$ which are affine in
momenta. In particular, the restriction to $V^*_tQ$ of any element
of the quantum algebra $\cA_V$ obeys this condition and,
consequently, belongs $\cA_t$. Conversely, any element of $\cA_t$
is of this type.$^{7,9}$  Thus, $\cA_t=i_t^*\cA_V$ and, therefore,
the polarization $\bT^*_V$ of the Poisson bundle $V^*Q\to\Bbb R$
is a fiberwise polarization.

In order to provide metaplectic correction and to complete
geometric quantization of symplectic fibers of the Poisson bundle
$V^*Q\to\Bbb R$, we use the fact that the Jacobian $J$ of the
transition function between bundle coordinates $(t,q^k)$ and
$(t,q'^k)$ on $Q$ coincides with the Jacobian $J_t$ of the
transition function between coordinates $(q^k)$ and $(q'^k)$ on
the fiber $Q_t$ of $Q$ over a point $t$. One can show that, given
a bundle  $\cD_{1/2}\to Q$ of complex half-forms on $Q$, its
pull-back $i^*_t\cD_{1/2}$ onto $Q_t$ is the fiber bundle of
half-forms on $Q_t$.$^{7,9}$ Then the formula (\ref{vv}) defines
the Schr\"odinger representation of the quantum algebra $\cA_t$ of
the symplectic fiber $Q_t$ by Hermitian operators (\ref{qq83'}) in
the pre-Hilbert space $E_t$ of half-forms x of compact support on
$Q_t$ equipped with the Hermitian metric (\ref{j29}). Spaces
$E_t$, $t\in \Bbb R$, are assembled into a trivial pre-Hilbert
bundle over $\Bbb R$.

Any half-form $\rho$ on $Q$ yields a half-form on $Q_t$. Given an
element $f\in \cA_V$ and its pull-back $f_t=i_t^*f\in \cA_t$, we
obtain from the formulae (\ref{qq83}) and (\ref{qq83'}) that $\wh
f\rho\circ i_t=\wh f_t(\rho\circ i_t)$. This equality shows that
the Schr\"odinger quantization of the Poisson manifold $V^*Q$ can
be seen as the instantwise quantization.


\end{document}